\documentclass[fleqn,twoside]{article}
\usepackage{espcrc2}

\usepackage{graphicx}
\usepackage[figuresright]{rotating}

\newcommand{\AmS}{{\protect\the\textfont2
  A\kern-.1667em\lower.5ex\hbox{M}\kern-.125emS}}
\title{Charged Higgs Flavor Changing Current in $\tau^- \to \nu_X K^- \pi^0$}
\author{D. Kimura\address[MCSD]{Learning Support Center, Hiroshima Shudo University, Hiroshima, 731-3195, Japan}
        ,Kang Young Lee\address[MCSD]{Division of Quantum Phases and Devices, School of Physics, Konkuk University, Seoul 143-701, Korea}
        ,T. Morozumi \address{Graduate School of Science,
 Hiroshima University,1-3-1,Higashi-Hiroshima,739-8526,Japan}and
        K. Nakagawa \addressmark}
\begin{document}
\def\endd{\end{document}}
\def\tbr{\textcolor{red}}
\def\tcr{\textcolor{red}}
\def\ov{\overline}
\def\nn{\nonumber}
\def\f{\frac}
\def\beq{\begin{equation}}
\def\eeq{\end{equation}}
\def\bea{\begin{eqnarray}}
\def\eea{\end{eqnarray}}
\def\ynu{y_{\nu}}
\def\ydu{y_{\triangle}}
\def\ynut{{y_{\nu}}^T}
\def\ynuv{y_{\nu}\frac{v}{\sqrt{2}}}
\def\ynuvt{{\ynut}\frac{v}{\sqrt{2}}}
\def\dc{\stackrel{\leftrightarrow}{\partial}}
\begin{abstract}
We study the effect of flavor changing charged current
(FCCC) of lepton sector in the hadronic tau decays,
$\tau \to \nu_X K^- \pi^0$ $(X=e,\mu)$.
In general two Higgs doublet model, the 
lepton flavor changing neutral current (FCNC) arises in 
the charged lepton sector and contributes to 
the process such as $\tau \to \mu \mu^+ \mu^-$
and $\tau \to \mu(e) P^0$ where $P^0=\pi^0,\eta, \eta'$.
We derive a relation between the FCCC
in charged Higgs boson and the FCNC due to the neutral Higgs boson.
Then by using the recent experimental upper bound on FCNC
of $\tau \to l_X \eta (\pi^0)$ ($l_X=e, \mu$)processes, 
we study how large the effect of FCCC could be in the process 
$\tau \to \nu_X K \pi$ decays. We also report the preliminary result
on the form factor calculations of $\tau \to K \pi \nu$ decays
and the hadronic invariant mass distribution.
\vspace{1pc}
\end{abstract}

\maketitle

\section{Introduction}
Thanks to the effort on search for lepton flavor violation
in B factories (Belle, Babar), the stringent constraints
on the upper bounds for branching ratios
for $\tau \to \mu P^0$ and  $\tau \to e P^0$ where
$P^0=\pi^0, \eta, \eta'$ are obtained. 
They constrain the Flavor Changing Neutral Current (FCNC)
couplings of charged lepton sector 
in various new physics models.
The examples of new physics models include super symmetric 
models, and multi-Higgs doublet models
\cite{Sher:02,Li:06,Chen:06}.
\section{Flavor Changing Charged Higgs interaction}
 In this talk, we take two Higgs doublet models as example. 
In type III two Higgs doublet model, 
there are tree level FCNC 
in charged lepton sector
which neutral Higgs bosons mediate.
CP odd Higgs boson ($A$) mediates the process like
\bea
\tau \to l_X A \to l_X \bar{q}{q} \quad, (l_X=e,\mu),
\label{eq:FCNC}
\eea 
where we focus on FCNC in lepton sectors.
In Eq.(\ref{eq:FCNC}) $\bar{q} q$ ($q=u,d,s$) 
forms pseudoscalar bilinear.  
At the same time, the charged Higgs boson mediates
the Flavor Changing Charged Current (FCCC) interaction.
\bea
&& \tau^- \to \nu_X H^- \to \nu_X 
K^- P^0 \quad, (X=e,\mu),
\label{eq:FCCC}
\eea
where 
$P^0=\pi^0, \eta, \eta^\prime$
and $H^-$ is the charged Higgs boson.
We study how large the contribution from
FCCC can be in the process of 
$\tau \to K \pi^0 \nu_X$ decay 
by considering the
experimental upper bounds on FCNC in charged lepton sector. 
In Table 1, we summarize
the experimental limits on FCNC from Belle and Babar
which are used in our study.
\begin{table}[htb]
\caption{The experimental upper limits on branching fractions
for $\tau \to l_X P^0$
decays:  The unit is $10^{-7}$.} 
\label{table:1}
\begin{center}
\begin{tabular}{|c|c|c|} \hline 
Process & Belle \cite{Miyazaki:07} & Babar
\cite{Aubert:07} \\ \hline
$\tau \to e \pi^0$ & $0.8$ & $1.4$  \\ \hline
$\tau \to \mu \pi^0$ & $1.2$ & $1.1$ \\ \hline
$\tau \to e \eta $ & $0.92$ & $1.9$\\ \hline
$\tau \to \mu \eta$ & $0.65 $ & $1.3$\\ \hline
$\tau \to e \eta^\prime$ & $1.6$ & $2.6$ \\ \hline
$\tau \to \mu \eta^\prime$ & $1.3$ & $2.0$ \\ \hline
\end{tabular}
\end{center}
\end{table}
In the
type III two Higgs doublet Model, both the
two vacuum expectation values
contribute to the mass of leptons ($m_l$), 
\bea
\tilde{H_1}&=&i \tau_2 H_1^{\ast}
=e^{-i \frac{\theta_{CP}}{2}}\left( \begin{array}{c}
-\sin \beta H^+ \\
-\frac{v_1+h_1+i \sin \beta A}{\sqrt{2}} \\
\end{array} \right),\nn \\
H_2&=&e^{i \frac{\theta_{CP}}{2}}\left( \begin{array}{c}
-\cos \beta H^+ \\
\frac{v_2+h_2-i \cos \beta A}{\sqrt{2}} \\
\end{array} \right).
\eea 
Because the Yukawa couplings for leptons are given by,  
\bea
-{\cal L}&=&y_{1 ij} \overline{e_{Ri}} \tilde{H_1}^{\dagger}
L_{Lj}+y_{2 ij} \overline{e_{Ri}} H_2^{\dagger} L_{Lj}
+ {\rm h.c.}, \nn \\
\eea
one obtains the charged lepton mass as,
\bea
m_l= V_R \frac{1}{\sqrt{2}} (-y_1 v_1 e^{i \frac{\theta_{CP}}{2}}
+y_2 v_2 e^{-i \frac{\theta_{CP}}{2}} ) V_L^{\dagger},
\label{chargedlepton}
\eea
where $V_R$ and $V_L$ are unitary matrices which diagonalize
the mass matrix for charged leptons. In general,
the neutral Higgs boson ($A, h_1, h_2$) couplings to leptons
are not flavor diagonal.
One can write the couplings in terms of the following 
dimensionless matrix $r_2$.
\bea
\frac{g}{\sqrt{2} M_W}
r_{2 ij} m_{lj}={\left(V_L y_2^\dagger V_R \right)}_{ij}
e^{i\frac{\theta_{\rm CP}}{2}}.
\eea
The Yukawa couplings
of CP odd Higgs boson to charged lepton and anti-lepton
are,
\bea
&&{\cal L}_{\rm NC}=- i A \frac{g}{2 M_W} \{
\tan \beta \overline{l_{i}} m_{li} \gamma_5
l_i \nn \\
&& +\frac{1}{\cos \beta}
 \overline{l_{i}} 
(m_{l} r_2^\dagger L -r_2  m_l R)_{ij} l_j\}
\eea
The FCNC couplings denoted by $r_{2 ij}$ for 
$l_j \to l_i A$   
are related
to FCCC couplings  $l_j \to \nu_i H^-$ as,
\bea
&&{\cal L}_{\rm FCCC}\nn \\
&&=-\frac{g} 
{\sqrt{2} M_W}H^+
\overline{\nu_{i}}
\{\delta_{ij} \tan \beta-\frac{r_{2ij}}{\cos \beta}\} m_{lj} 
R l_{j}.\nn \\
\eea
\section{Constraints on FCNC couplings from 
$\tau \to \mu (e) P^0$ decays}
It is straightforward to obtain the constraints on FCNC
couplings from the charged lepton FCNC processes
\cite{Li:06}.
\bea
&& Br(\tau \to l_X \pi^0)=
\frac{p_{\pi^0}}{8 \pi \Gamma_\tau} \times  \nn \\
&& \left(\frac{f m^2_{\pi^+} m_{\tau} G_F}{\sqrt{2} M_A^2 \cos^2 \beta}
\right)^2
(\tan \beta \Delta_d -\cot \beta \Delta_u)^2 \nn \\
&&\{ \frac{ 1+ \delta^2_{l_X}-\delta^2_{\pi^0} }{2}
(|r_{2 X \tau}|^2 + |r_{2 \tau X}|^2 \delta^2_{l_X}),
\nn \\
&&-\delta^2_{l_X}(r_{2 \tau X} r_{2 X \tau}+ h.c.) \}.
\eea
where $\delta_{lX}= \frac{m_{l_X}}{m_\tau}$,
$f$ is pion decay constant and 
$\Delta_{d(u)}=\frac{2 m_{d(u)}}{m_u+m_d}$.
We require that the predictions are smaller than experimental
upper limits; $
Br(\tau \to l_X \pi^0) \le Br^{\rm UL}_{\rm exp.}$.
We show the constraints on $(|r_{2 X \tau}|,|r_{2 \tau X}|)$ 
plane in Fig.1
\begin{figure}
\begin{center}
\includegraphics[width=6cm]{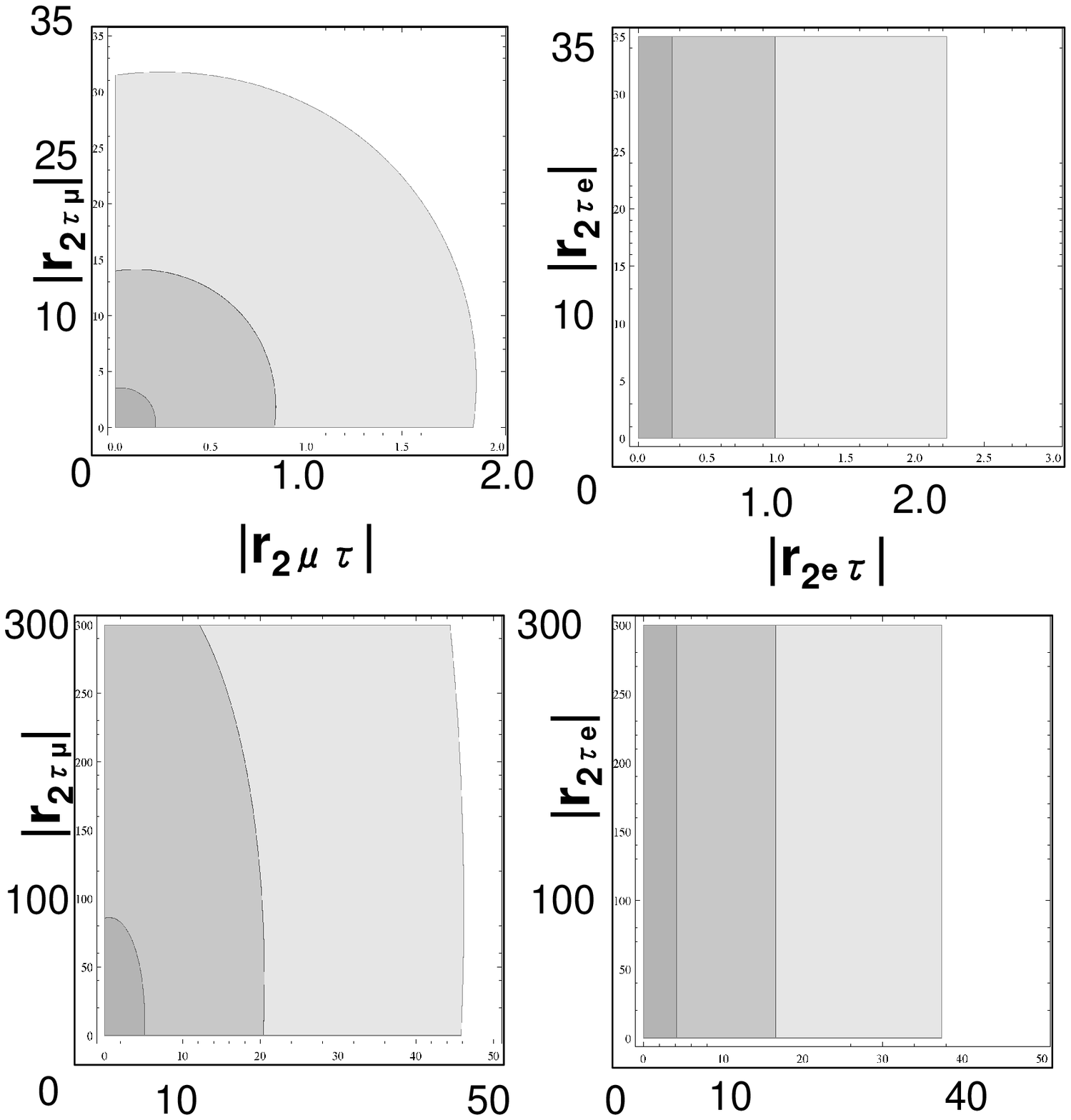}
\end{center}
\caption{Upper bounds on FCNC couplings 
$(|r_{2 X \tau}|,|r_{2 \tau X}|), \quad X=e, \mu$.  
The top two panels obtained from the decays 
$\tau \to l_X  \eta$ and
the bottom two panels from $\tau \to l_X \pi^0$. 
The dark shaded regions correspond to $M_A=100$ GeV and the 
brightest ones correspond to $M_A=300$ GeV.
The region in the middle of them corresponds to $M_A=200$ GeV.
 $\tan \beta$ is chosen as $14.1$.}
\end{figure}
\section{Summary of constraints}
We summarize the constraints on FCNC couplings.
\bea
|r_{2 \mu \tau}| <0.9 (\frac{M_A}{200})^2,
|r_{2 e \tau}|<1 (\frac{M_A}{200})^2.
\eea
The constraints from $\tau \to \eta$ mode are
more stringent than those of $\tau \to \pi^0$ because 
the coupling of the strangeness quark 
with  Higgs
is much stronger than  those of the other light quarks,
\bea
&& Br(\tau \to l_X \eta ) 
\ge \frac{p_{\eta}}{24 \pi \Gamma_\tau} 
\left(\frac{f m^2_{\pi^+} m_{\tau} G_F}{\sqrt{2} M_A^2 \cos^2 \beta}
\right)^2  \nn \\
&& (\tan \beta (\Delta_d-2 \Delta_s)
+\cot \beta \Delta_u)^2  \nn \\
&& \{ \frac{ 1+ \delta^2_{l_X}-\delta^2_{\eta} }{2}
(|r_{2 X \tau}|^2 + |r_{2 \tau X}|^2 \delta^2_{l_X}) \nn \\
&&-2\delta^2_{l_X}|r_{2 \tau X} r_{2 X \tau}|) 
\},
\eea
where $\Delta_s=\frac{2m_s}{m_u+m_d}
\sim 26, \Delta_d \sim 1.5 , \Delta_u \sim 0.5$
and we treat $\eta$ meson as a pure octet.
\section{Possible effect on FCCC process}
Using the constraint determined 
by FCNC processes and assuming charged Higgs boson mass
, one can estimate how large FCCC can be.
The hadronic invariant mass ($\sqrt{s}$) distribution for $\tau 
\to K^- \pi^0 \nu $ decay including the FCCC effect is 
\bea
&&\sum_{X=e, \mu, \tau}
\frac{d{\rm Br}(\tau \to \nu_X K^- \pi^0)}{d\sqrt{s}}=\nn \\
&&\frac{1}{\Gamma} 
\frac{G_F^2 |V_{us}|^2}{2^5 \pi^3} 
\frac{(m_{\tau}^2-s)^2}{m_{\tau}^3} p_{K} \nn \\
&& \left(\frac{m_{\tau}^2}{2} \Large|1-\frac{s}{M_H^2} \tan^2 \beta(1-
\frac{r_{2 \tau \tau}}{\sin \beta})\Large|^2
|F_s|^2 \right. \nn \\
&& \left. +\frac{m_{\tau}^2}{2}
(\frac{s \tan^2 \beta}{M_H^2 \sin \beta})^2
(|r_{2 e \tau}|^2+|r_{2 \mu \tau}|^2)
|F_s|^2 \right),\nn \\
&& \left.+(\frac{2 m_{\tau}^2}{3 s}+\frac{4}{3}){p_{K}}^2|F|^2
\right),
\eea
where $F$ is vector form factor and $F_s$ is scalar form factor
defined by,
\bea
&& \langle \pi^0 K^+|\bar{u} \gamma_\mu s|0 \rangle =\nn \\
&&F(Q^2) q_\mu
+(F_s(Q^2)-\frac{\Delta_{K \pi}}{Q^2} F(Q^2) ) Q_\mu,
\label{eq:ff}
\eea
and
$Q_\mu=p_K+ p_\pi, q=p_K-p_\pi$.
In Fig.2, We show 
the FCCC contribution normalized by the standard model
contribution,
\bea
R(\sqrt{s})=\frac{\sum_{X=e, \mu}\frac{d B_r(\tau \to K^- \pi^0 \nu_X)}{d \sqrt{s}}}
{\frac{d B_r(\tau \to K^- \pi^0 \nu_\tau)}{d \sqrt{s}}{\Huge|}_{\rm S.M.}}. 
\label{eq:Ratio}
\eea
We found the effect of FCCC is at most  
O$(10^{-7}) \sim O(10^{-4})$ for the charged Higgs mass 
$M_H=215$(GeV) when the flavor changing couplings are chosen
as 
$r_{e \tau}=0.9,
r_{\mu \tau}=1$ and $\tan \beta$ is $14.1$.
\begin{figure}
\begin{center}
\includegraphics[width=7cm]{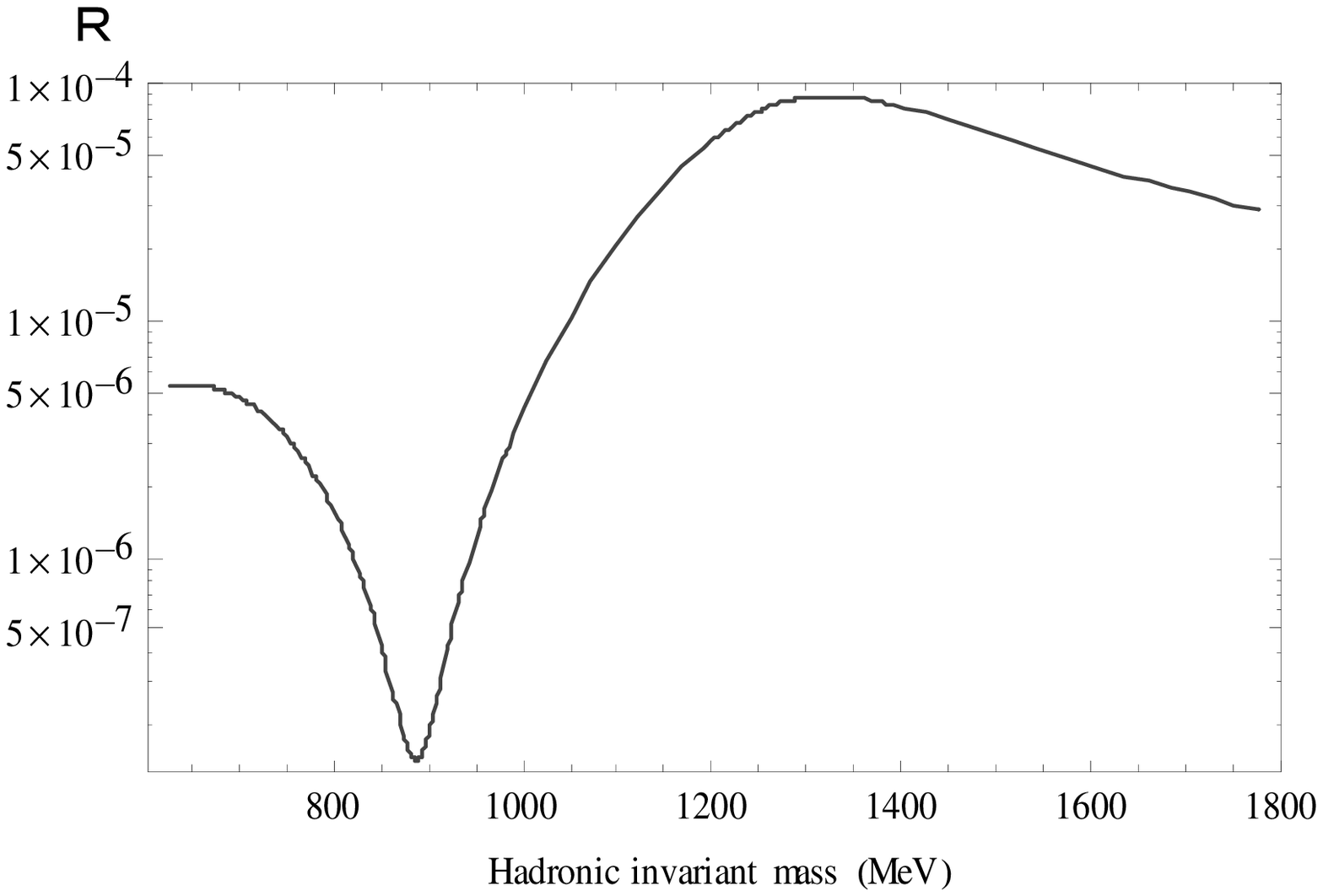}
\end{center}
\caption{The FCCC contributions to the hadronic mass
distribution in $\tau \to \nu_X K \pi$ decays ($X=e, \mu$).}
\label{fig:ratio}
\end{figure}
\section{The form factors of $\tau \to K \pi \nu$}
To obtain  Fig.\ref{fig:ratio}, the form factors
for the process have been computed.
Using the form factors in Eq.(\ref{eq:ff}), one can compare the 
theoretical predictions with the experimentally measured
one. They are important for study of the hadronic mass distribution
and the angular distribution of $\tau \to K \pi \nu$ decay \cite{Kimura}.
In experimental side, 
Belle and Babar measured the hadronic mass distribution
\cite{Belle:07,Babar:07} 
for $\tau^+ \to K_s \pi^+ \overline{\nu}$ and
$\tau^- \to K^- \pi^0 \nu$ respectively. The former
process is identical to
$\tau^+ \to K^+ \pi^0 \overline{\nu}$ in the isospin limit
with CP violation of $K^0 \overline{K^0}$ mixing 
neglected. 
We summarize the results related to $\tau^{\pm} \to K_s \pi^{\pm} 
\nu$ \cite{Belle:07}.
\begin{itemize}
\item The hadronic mass distribution and the branching 
fraction are measured. 
$Br( \tau^- \to K_s \pi^- \nu)=0.404\pm 0.002 \pm 0.013\%$
\item 
The experimentally measured
spectrum is combined with the Breit-Wigner form of the
several resonances.  The fit to the spectrum 
leads to $K^\ast$ mass larger than the mass measured in 
hadronic processes.  Belle's result leads to
$m_{K^\ast}=895.47\pm0.20 \pm 0.44 \pm 0.59$ MeV, which is 
larger than  $891.66\pm0.26$ MeV.  The latter value
is hadronically produced
$K^\ast$ mass average of PDG 2010.\\
The width of $K^\ast$ is measured by the same process. The obtained value 
$\Gamma_{K^\ast}=46.2\pm0.6 \pm 1.0\pm 0.6 \pm 0.7$ MeV is also different from
the width measured in hadronic reactions which average 
is $50.8 \pm 0.9$ MeV.
\item  Belle estimated the branching fraction 
for the process going through the $K^\ast$
resonance as,
$Br(\tau^- \to K^{\ast-} \nu)Br(K^{\ast -} \to K^- \pi^0)=
(3.77\pm0.02\pm0.12\pm0.12) \times 10^{-3}$. Combined it with
$Br(K^{\ast-} \to K_s \pi^-)=\frac{1}{3}$,
PDG 2010 estimated $Br(\tau^- \to K^{\ast -} \nu) \sim 1.13 \%$. 
\end{itemize}
\section{Chiral Lagrangian including vector resonance}
To compute the form factors, we use the chiral Lagrangian
including vector resonances. A new point of our analysis
is that we compute the corrections 
in one loop level without spoiling
the chiral counting. The rigorous one loop amplitude
is identical to the one of chiral perturbation.
However, in order to reproduce the effect of intermediate
resonance contribution, one needs to go beyond the 
simple O$(p^4)$ chiral perturbation. For the purpose,
we resum the chiral corrections to vector meson self-energy.
We perform the resummation so that the procedure does not
spoil the O$(p^4)$ corrections. Therefore our method is consistent
with the chiral perturbation at O($p^4$) level. This feature is 
important so that our frame work reproduces the correct behaviour
for the hadronic invariant mass distribution at threshold
where chiral perturbation is valid. On the other hand, by resumming
the self-energy corrections, one can recover the resonance 
behaviour even far away from the threshold region.
To renormalize the one loop amplitudes, we identify the
counter terms.

Denoting
$V$ as SU(3)$_f$ octet of vector mesons ($K^\ast, \rho, ...$)
and $\pi$ as octet pseudoscalars ($\pi, K, \eta_8$),
the Lagrangian including the counter terms are given as,
\bea
{\cal L}={\cal L}^{CHPT (0)}+M_{V}^2 {\rm Tr}
(V_{\mu}-\frac{\alpha_{\mu}}{g})^2
+
{\cal L}_{c},
\eea
where 
$
\alpha_\mu \sim \frac{1}{f^2} [\pi, \partial_\mu \pi].
$
$M_V$ denotes the chiral limit mass for vector meson.
In principle,
the value can be extracted from the lattice calculation by
extrapolating the vector meson mass with the finite pseudoscalar
meson mass to the chiral limit, i.e.,
$m_u,m_d,m_s \to 0$. 
As for the counter terms, we note the following 
counter terms are needed to subtract the divergences in
one loop level.
\bea
{\cal L}_c&=&{\cal L}^{CHPT (2)}-\frac{Z_V}{2}
{\rm Tr} F_{V \mu \nu} F_V^{\mu \nu}
\nn \\
&+&C_1 {\rm Tr}\left(\frac{\xi \chi \xi + \xi^{\dagger} 
\chi^{\dagger}
 \xi^{\dagger}}{2}\right)(V_\mu-\frac{\alpha_{\mu}}{g})^2 
\nn \\
&+&C_2 {\rm Tr}\left(\frac{\xi \chi \xi + \xi^{\dagger} \chi^{\dagger}
 \xi^{\dagger}}{2}\right)
 {\rm Tr}(V_{\mu}-\frac{\alpha_{\mu}}{g})^2\nn \\
&+& 
i C_3  {\rm Tr} F_V^{\mu \nu} 
\alpha_{\perp \mu} \alpha_{\perp \nu} \nn \\
&+&C_4 {\rm Tr}(\xi F_V^{\mu \nu} \xi^{\dagger} F_{L \mu \nu})
\quad, \alpha_{\perp} \sim \frac{\partial \pi}{f},
\label{eq:cv}
\eea
where $ \chi={\rm diag.}
(m_\pi^2, m_\pi^2, 2 m_K^2-m_\pi^2) 
$ and $\xi=e^{i \frac{\pi}{f}}$.
$C_1 \sim C_4$ and $Z_V$ are coefficients of the counter terms. 
\section{The strangeness changing
charged  current in terms of hadrons}
In the model in Eq.(\ref{eq:cv}), the strangeness changing charged current is given as, 
\bea
&& \overline{u_L} \gamma_\mu s_L= \frac{M_V^2}{\sqrt{2} g} K_\mu^{\ast -}\nn \\
&& -i\frac{K^- \dc \pi^0}{2 \sqrt{2}} 
\left( (1-\frac{M_V^2}{2 g^2 f^2})
\sqrt{Z_K Z_\pi} \right. \nn 
\\
&&\left.+
\frac{2 m_K^2+ m_\pi^2}{f^2}(8L_4-\frac{C_2}{2g^2}) \right.\nn \\
&&+ \left.
\frac{2 m_K^2}{f^2}(4 L_5 -\frac{C_1}{2 g^2}) \right) 
+\frac{4i L_5}{\sqrt{2}f^2}\Delta_{K \pi}
K^- \partial_\mu \pi^0  \nn \\
&&+ \frac{\sqrt{2} i L_9}{f^2} \partial^\nu
(\partial_\nu K^- \partial_\mu \pi^0-\partial_\mu K^-
\partial_\nu \pi^0).
\eea
Within tree level, 
including the contribution of $K^*$ exchange diagram,
the result of the low energy theorem
is reproduced as the sum of the direct $K \pi$ production
process and 
the $K^*$ production process as the intermediate state,
\bea
&&<\pi^0 K^+|\bar{u}\gamma_\mu s|0>|_{\rm dir.}=-\frac{q_\mu}{\sqrt{2}}
(1-\frac{M_V^2}{2 g^2 f^2}), \nn \\
&& \\
&&<\pi^0 K^+|\bar{u}\gamma_\mu s|0>|_{\rm K^*}=
\frac{M_V^2 q_\mu}{4g f^2}
\frac{-1}{M_V^2} \frac{\sqrt{2} M_V^2}{g}. \nn \\
\eea
The sum of them leads to,
\bea
&& F= -\frac{1}{\sqrt{2}}, \quad F_s=\frac{\Delta_{K \pi}}{Q^2} F.
\eea
Beyond the tree level, one needs to
compute
four topologies of Feynman diagrams.
\begin{figure}
\begin{center}
\begin{tabular}{c} 
\includegraphics[width=5.0cm]{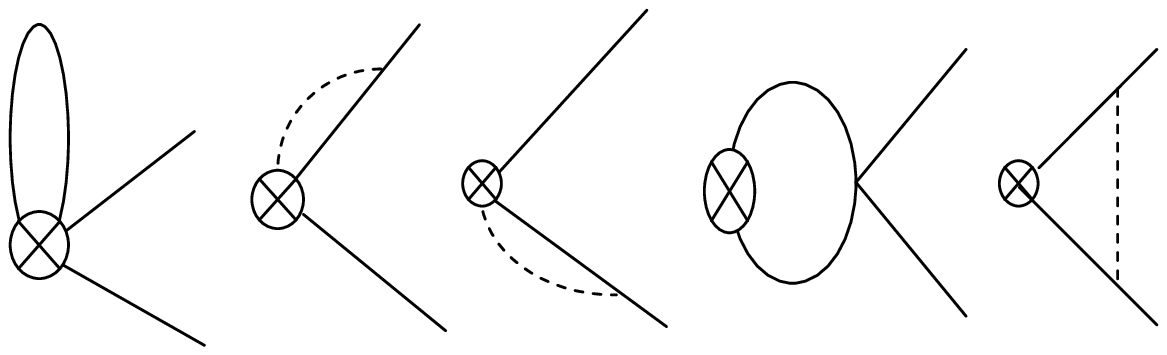}
\\ 
\includegraphics[width=5.0cm]{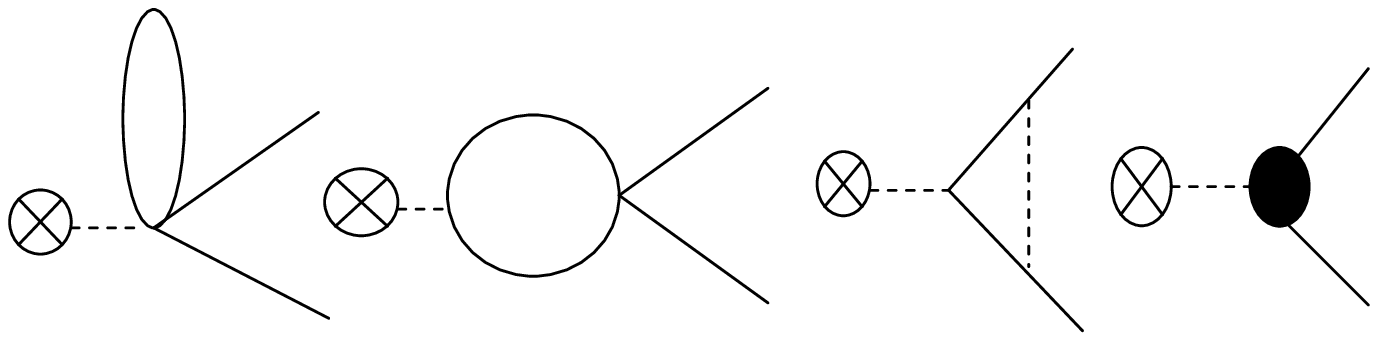}
\\ 
\includegraphics[width=5.0cm]{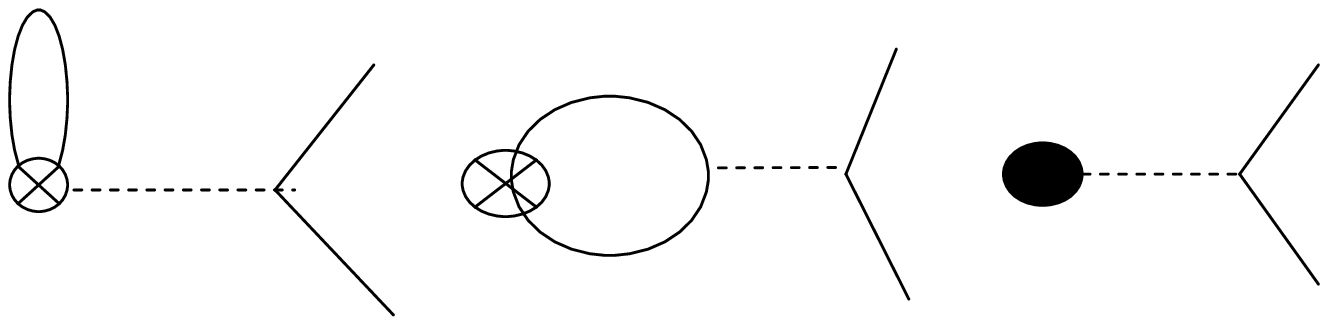}
\\ 
\end{tabular}
\end{center}
\caption{The one loop diagrams which contribute
to $K \pi$ form factors. The crossed circle
($\otimes$)
denotes the insertion of the charged current
.
The solid lines (the dotted lines)
correspond to pseudoscalar(vector) meson. The vector meson
propagator
is $\frac{i g_{\mu \nu}}{M_V^2}$. 
The black blob denotes the counter term.
In the top line,  one loop corrections
to $K \pi$ vertex are shown. 
In the second line, the diagrams with one loop
corrections to $K^{\ast} \to K \pi$ vertex are shown. 
In the bottom line, the diagrams with one loop  
corrections to the $K^\ast$ production amplitude
are shown.} 
\label{fig:1}
\end{figure}
The form factors in one-loop level correspond to the
Feynman diagrams shown in Fig.3 and the self-energy
diagrams shown in Fig.4.
By adding the tree and one loop amplitudes, we found the
result is the same as that of the one-loop chiral perturbation theory
\cite{GasserI:86} except the point that 
one needs to replace the counter term $l_9$ with $l_{9 \rm{eff}}$,
\bea
l_{9 \rm{eff}}=l_9+\frac{c_3}{8g}-\frac{c_4}{2g}+\frac{z_V}{4g^2},
\eea
where $l_9$,$c_3$, $c_4$ and $z_V$ are finite parts of the
coefficient of the counter terms. Then the vector
form factor can be written with the function,
\bea
&&f_+^{CHPT}(s)=1+\frac{3}{2}(H_{K \pi}(s)+H_{K \eta}(s)),\\
&&H_{PQ}=\frac{1}{f^2} (s M_{PQ}(s) -L_{PQ}(s)) + \frac{2}{3f^2}
l_{9 {\rm eff}} s, \nn
\eea
where the functions $H,M,L$ are defined in \cite{GasserI:86}.
\subsection{Form Factors beyond one loop}
In our framework, we can resum the self-energy diagrams
of the vector meson. The resummed propagator may have
the pole at complex plane which corresponds to 
the resonance ($\rho, K^\ast$).
\begin{figure}[htbp]
\resizebox{5cm}{!}{\includegraphics{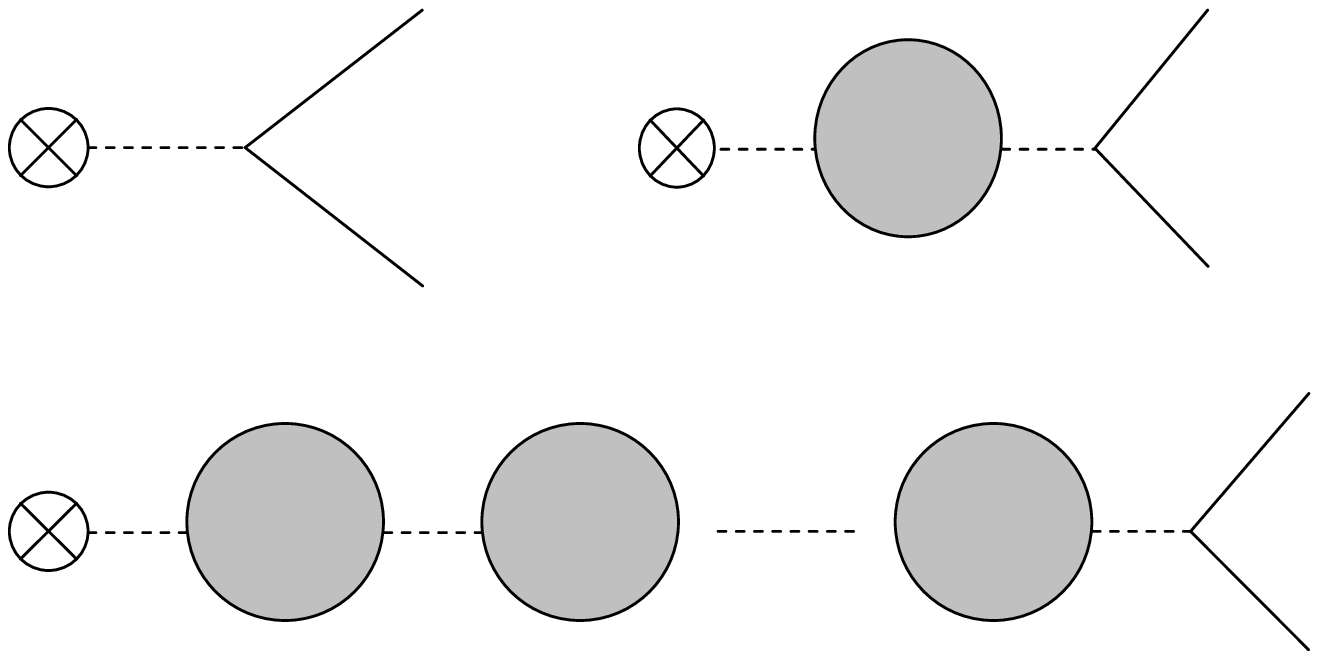}}
\caption{The form factor contribution from
the self-energy of $K^*$. The self-energy corrections 
due to $K,\pi,\eta$ mesons loop
are summed as the dressed propagator.}
\label{fig:fig5}
\end{figure}
The resummed propagator is obtained by inverting the
one loop corrected inverse propagator.
Denotaing $\delta A$ and $\delta B$ as chiral corrections
to vector meson self energy including the counter terms
of the Lagrangian $(Z_V,C_1,C_2)$, 
the relation between the propagator and its inverse is,
\bea
[(M_V^2 + \delta A)g^{\mu \nu}+ \delta B Q^{\mu} Q^{\nu})]
D_{\nu \rho}= \delta^{\mu}_{\rho}.
\eea
Then the resummed propagator can be easily obtained.
\bea
D_{\mu \nu}=\frac{1}{M_V^2+ \delta A}(g_{\mu \nu}-
\frac{Q_\mu Q_\nu \delta B}{M_V^2+ \delta A + Q^2 \delta B}).
\eea
The new contribution due to the resummed propagator
to the form factor is given as: 
\bea
&& \delta <\pi^0 K^+|\bar{u}\gamma_\mu s|0>
= -\frac{M_V^4}{2 \sqrt{2} g^2 f^2}q^{\nu} \nn \\
&& (D_{\nu \mu}-\frac{g_{\nu \mu}}{M_V^2} + 
\frac{\delta A g_{\nu \mu} +Q_\nu Q_\mu \delta B}{M_V^4}).
\eea
Note that we subtracted the contribution due to
the bare propagator and once self-energy insertion.
Then the form factors for which the resummation is taken account 
of are given as,
\bea
&& F_s=-\frac{\Delta_{K \pi}}{\sqrt{2} Q^2} \times  \nn \\
&& \left(f_0^{\rm CHPT}
+\frac{(\delta A_K^\ast + Q^2 \delta B_K^\ast)^2}
{2 g^2 f^2 (M_V^2 + \delta A_K^\ast + Q^2 \delta B_K^\ast)}
\right), \nn \\
&& F=\frac{-1}{\sqrt{2}}
\left(f_+^{\rm CHPT}
+\frac{1}{2 g^2 f^2}  
\frac{(\delta A_K^\ast)^2}{M_V^2 + \delta A_K^\ast} \right).
\eea
\section{Comparison with the Belle Data}
The prediction on hadron invariant mass spectrum
using our form factors
is compared with 
the data of $\tau \to K_s \pi^- \nu $
from \cite{Belle:07} shown with error 
bars in Fig.5.
\begin{figure}
\begin{center}
\includegraphics[width=8cm]{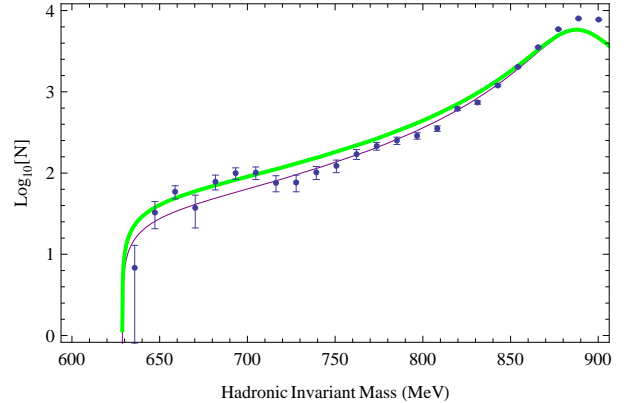} 
\end{center}
\caption{The hadronic invariant mass
spectrum for $\tau \to K \pi \nu$ 
decay at low invariant mass region. 
The green thick solid line and the thin purple line correspond
to predictions based on
the theoretical model with 
$M_V=700$ and $850$ (MeV), respectively.
The data is taken from \cite{Belle:07}.}
\end{figure}
At the low invariant mass region, our result is consistent
with the measured spectrum. However, around the resonance region, our result is smaller than the measured spectrum.
The improved treatment will be given elsewhere. 
A previous study with the dispersive approach is shown in 
\cite{JaminI:08}.
\section{Acknowledgement}
We would like to thank Prof. George Lafferty and his colleagues
for organizing the workshop. We also thank for 
Dr.D. Epifanov for providing us with
the hadronic mass distribution.  
KYL is supported in part by WCU program through the KOSEF funded
by the MEST (R31-2008-000-10057-0)
and the Basic Science Research Program through the National Research Foundation
of Korea (NRF) funded by the Korean Ministry of
Education, Science and Technology (2010-0010916).
TM is supported by Grant-in-Aid for Scientific Research (C)
(No.22540283) from JSPS.

\end{document}